\documentclass{emulateapj}
\usepackage{graphicx}
\usepackage{epsfig}
\usepackage{rotating}

\newcommand{\bc}{\begin{center}}
\newcommand{\ec}{\end{center}}
\newcommand{\be}{\begin{equation}}
\newcommand{\ee}{\end{equation}}
\newcommand{\ba}{\begin{eqnarray}}
\newcommand{\ea}{\end{eqnarray}}

\def\farcs{\hbox{$.\!\!^{\prime\prime}$}}

\def\chan{{\sl Chandra}}
\def\xmm{{\sl XMM-Newton}}

\begin{document}

\submitted{\today}

\title{
X-ray emission from the double neutron star binary B1534+12: Powered
by the pulsar wind?}

\author{
 O.\ Kargaltsev, G.\ G.\ Pavlov, and G.\ P.\ Garmire}
\affil{The Pennsylvania State University, 525 Davey Lab, University
Park, PA 16802, USA} \email{pavlov@astro.psu.edu}

\begin{abstract}
We report the detection of the double neutron star binary (DNSB)
B1534+12 (= J1537+1155) with the \chan\ X-ray Observatory. This DNSB
($P_{\rm orb} = 10.1$ hr) consists of the millisecond (recycled)
pulsar J1537+1155A ($P_A=37.9$ ms) and a neutron star not detected
in the radio. After the remarkable double pulsar binary
J0737$-$3039, it is the only other DNSB detected in X-rays. We
measured the flux of $(2.2\pm 0.6)\times10^{-15}$ ergs s$^{-1}$
cm$^{-2}$  in the 0.3--6 keV band. The small number of collected
counts allows only crude estimates of spectral parameters. The
power-law fit yields the photon index $\Gamma = 3.2\pm 0.5$ and the
unabsorbed 0.2--10 keV luminosity $L_{\rm X}\approx6\times10^{29}\,
{\rm ergs}\, {\rm s}^{-1} \approx 3\times 10^{-4} \dot{E}_A$, where
$\dot{E}_A$ is the spin-down power of J1537+1155A. Alternatively,
the spectrum can be fitted by a blackbody model with $T\approx 2.2$
MK and the projected emitting area of $\sim 5\times 10^3$ m$^2$. The
distribution of photon arrival times over binary orbital phase shows
a deficit of X-ray emission around apastron, which suggests that the
emission is caused by interaction of the relativistic wind from
J1537+1155A with its neutron star companion. We also reanalyzed the
{\sl Chandra} and {\sl XMM-Newton} observations of J0737$-$3039 and
found that its X-ray spectrum is similar to the spectrum of
B1534+12, and its X-ray luminosity is about the same fraction of
$\dot{E}_A$, which suggests similar X-ray emission mechanisms.
However, the X-ray emission from J0737$-$3039 does not show orbital
phase dependence. This difference can be explained by the smaller
eccentricity of J0737$-$3039 or a smaller misalignment between the
equatorial plane of the millisecond pulsar and the orbital plane of
the binary.
\end{abstract}
\section{Introduction}

 The sample of known double neutron star binaries (DNSBs) has nearly doubled over the
 last six years. Currently, eight such binaries are known
(they are shown in the $P$-$\dot{P}$ diagram
 in Fig.\ 1). All of them are found
 via radio timing of
  pulsars which provides
binary orbital
 parameters, including the companion masses. DNSBs are thought to be
 formed in consequent supernova explosions in the course of evolution
 of massive binary
 systems.
  The details of binary evolution depend on
 initial star masses and binary orbital parameters (see Stairs 2004
for a review).
  It is generally believed
that once the first neutron star (NS) is formed in the binary,
accretion of matter from the evolved second star
spins up the NS so that it becomes a
millisecond (recycled) pulsar (MSP).
Eventually, the second
star explodes as a supernova and, if the explosion does not disrupt
the binary or disintegrate the second star,
a DNSB (or a NS-black hole binary) is
formed.

Radio observations of DNSBs allow one to measure the masses of both
NSs and test the predictions of General Relativity (GR). The recent
discovery of the tight DNSB J0737$-$3039 ($P_{\rm orb}=2.45$ hours;
see Burgay et al.\ 2003; Lyne et al.\ 2004; and Tables 1 and 2 for
other details), where both NSs are radio pulsars orbiting in the
plane seen nearly edge-on (the inclination angle $i$ differs from
$90^\circ$ by $0\fdg 29 \pm 0\fdg 14$; Coles et al.\ 2005),
dramatically boosted interest in DNSBs. Apart from measuring the GR
parameters, the radio observations have  shown variations of the
pulsed flux with orbital phase, for both pulsars.
 The radio pulse of the MSP J0737$-$3039A
(hereafter J0737A) disappears around its superior
  conjunction\footnote{
At the superior (inferior) conjunction of J0737A, it is located
behind (in front of) J0737B along the observer's
 line-of-sight.},
which has been interpreted as an eclipse of J0737A ( Lyne et al.\
2004), caused by synchrotron absorption in the magnetosphere of
J0737B (Rafikov \& Goldreich 2005; Lyutikov \& Thompson 2005).
 Even more interesting is
the finding that the generally faint pulsed radio emission of J0737B
is strongly enhanced during two short orbital phase intervals of
$\sim 0.1$ of the period around the phase of inferior conjunction of
J0737B. These changes suggest that the magnetosphere of J0737B
responds to the wind or radiation of the more energetic J0737A
 (Lyne et
al.\ 2004;
Jenet \& Ransom 2004;
Demorest at
al.\ 2004;
Zhang \& Loeb 2004).

 The J0737 was
 the first and, until this work, the only DNSB detected in
 X-rays.
 It has been observed with both \chan\ (McLaughlin et al.\ 2004) and
 \xmm\
 (Campana, Possenti \& Burgay 2004; Pellizzoni et al.\ 2004).
Its luminosity
  is $L_{\rm X}\sim (2$--$3) \times 10^{30}$ ergs s$^{-1}$ in the 0.2--10 keV
 band,
  assuming a 500 pc distance. The X-ray spectrum is rather soft;
it can be fitted
by  either
 a power-law (PL) model with a photon index
$\Gamma \sim 3$--4
or a blackbody (BB)
 model with
$kT\approx 0.2$ keV and projected emitting area
$A\sim 3\times 10^4$ m$^2$.
 No X-ray variability has been reported for J0737 so far.
A number of different
 interpretations of the X-ray emission
 have been suggested. These include emission from
the surface and/or magnetosphere of J0737A (similar to
solitary
 MSPs),
emission from the J0737B's magnetosphere (surface) energized (heated)
by the J0737A's wind (Zhang \& Loeb 2004),
 and emission by particles accelerated in
a bow shock
produced by the J0737A's wind in the interstellar medium (ISM)
or in the magnetosphere of J0737B  (Granot \& M\'esz\'aros 2004).
  However, because of
the small number of X-ray counts collected, it is difficult to
discriminate between these models from observations of just one
DNSB.

In this paper we report the detection of X-rays from
the second DNSB, B1534+12 (=J1537+1155; hereafter
 J1537), discovered by Wolszczan (1991).
One of the two NSs in this DNSB is a millisecond pulsar
(J1537A hereafter; period $P_{A}=37.9$ ms,
  characteristic age $\tau_{A}\equiv P_{A}/2\dot{P}_{A}=248$
 Myr, spin-down power
 $\dot{E}_{A}=1.8\times10^{33}$ ergs s$^{-1}$)  on a 10.1 hour
 eccentric ($e=0.274$) orbit.
 As one can see from Figure 1 and Table 2,
the J1537A's parameters are generally close to those of
 J0737A,
although J1537A is a factor of 3 less energetic.
  The dispersion measure of J1537A suggests a distance of about 0.7 kpc.
The comparison of the observed orbital period decay
 with that predicted by GR
 allowed Stairs et al.\ (2002)
to evaluate the distance more accurately: $D=1.02\pm0.05$ kpc.
 Multi-year timing observations in the radio (Stairs et al.\ 2004)
suggest that we are looking at J1537 close to
 its orbital plane ($i\simeq 77\fdg 2$), and J1537A is
an almost orthogonal rotator (i.e., its magnetic axis and
the line of sight are nearly orthogonal to
the spin axis:  $\alpha\approx103^{\circ}$,
$\zeta\approx 97^\circ$ at the epoch of our observation\footnote{The angle
$\zeta$ is changing with time, $d\zeta/dt \simeq -0.21^\circ /{\rm yr}$
as a result of geodetic precession.}) whose spin axis is
 misaligned with the orbital angular momentum by
$\delta = 25^{\circ}\pm 4^\circ$
(Stairs et al.\ 2004; Thorsett et al.\ 2005).
The second NS (hereafter J1537B)
has not been detected
as a radio pulsar,
but its mass, $M_{B}=(1.345\pm0.001)M_{\odot}$, has been
measured (Stairs et al.\ 2002). The detection of X-rays from J0737
prompted us to observe J1537, the second nearest DNSB.
 We describe the {\sl Chandra} observation of J1537 in \S2 and
the observational results in \S3, which also includes our reanalysis
of the X-ray observations of J0737 and a comparison of the X-ray
properties of these DNSBs. We discuss plausible X-ray emission
models and compare their predictions with the observed X-ray
properties in \S4 and summarize our results in \S5.

\section{Observation}

We observed J1537 with the Advanced CCD Imaging Spectrometer (ACIS)
aboard {\sl Chandra} on 2005 April 10 (start time
53,470.352222 MJD).
 The useful scientific exposure time was 36,080 s
(first 935 s of the 37,015 s total exposure
 are excluded from the Good Time Intervals [GTIs]). The
observation was carried out in Very Faint mode, and the pulsar was
imaged on S3 chip of the ACIS-S array. The detector was operated in
Full Frame mode which provides time resolution of 3.2 seconds. The
data were reduced using the Chandra Interactive Analysis of
Observations (CIAO) software (ver.\ 3.2.1; CALDB ver.\ 3.0.3).

\begin{figure}
 % \center
 \vspace{-1.0cm}
\includegraphics[width=2.8in,angle=90]{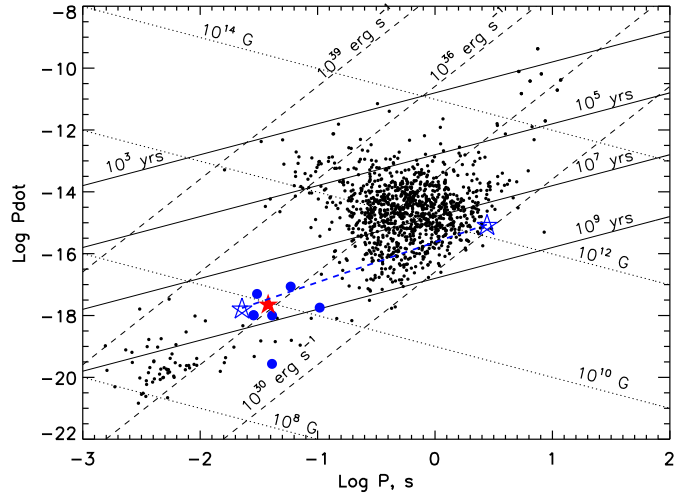}
\caption{ $P$-$\dot{P}$ diagram for $\approx1400$ radio pulsars
(small dots) from the ATNF catalog (Manchester et al.\ 2005). Lines
of constant pulsar age, magnetic field, and $\dot{E}$ are shown.
Eight DNSBs are marked by different symbols. J1537A is marked by the
filled star. The J0737 components are marked by open stars connected
with the straight dashed line. The rest of the DNSBs are marked by
filled circles. {\em See the electronic edition of the Journal for
the color version of this figure.} }
\end{figure}

\section{X-ray image and spectrum}

Figure 2 shows the ACIS-S3 image of the J1537's field. The X-ray
source is clearly seen
at ${\rm R.A.}=15^{\rm h}37^{\rm
m}09\fs986$, ${\rm Decl.}=+11^{\circ}55' 55\farcs39$
(the combined aspect determination and centroiding uncertainty
is 0\farcs7 at a 90\% confidence level).
Since this position is within $0\farcs5$ of the J1537A's radio
position
(Stairs et al.\ 2002),
we conclude with confidence that we detected the X-ray
emission from J1537. The distribution of source counts in the
ACIS image is consistent with that of a point source.

\begin{figure}[t]
 \centering
\includegraphics[width=2.6in,angle=0]{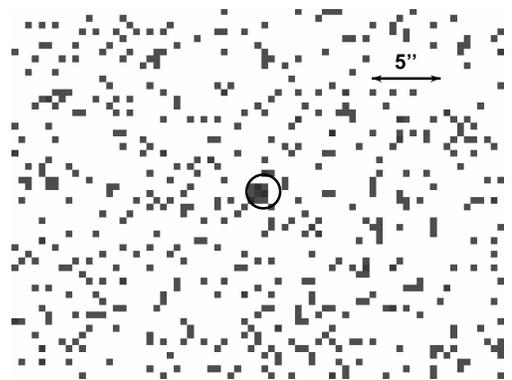}
\caption{ ACIS-S3 image of the J1537 field. The circle of 1\farcs23
radius shows the radio position of PSR J1537+1155 from Stairs et
al.\ (2002). }
\end{figure}

We extracted the pulsar spectrum from a circular aperture of
2.5 ACIS pixels radius ($=1\farcs23$; $\approx90$\% encircled energy radius)
 using the CIAO {\tt psextract} task. The background
was extracted from a $5''<r<22''$ annulus centered on the source.
The total number of counts within the source aperture is 16, of
which 98.4\% are expected to come from the source. The observed
source flux is $(2.2\pm0.6)\times10^{-15}$ erg s$^{-1}$ cm$^{-2}$ in
the 0.3--6 keV range.
 We group the counts
into 4 spectral bins, with each bin having 3--5 counts (see Fig.\
3), and fit the spectrum with absorbed BB and PL models using the
C-statistic (Cash 1979) implemented in XSPEC, ver.\ 11.3.0 (the
fitting parameters are given in Table 3). To obtain constrained
fits, we had to freeze the hydrogen column density, $N_{\rm H}$. The
pulsar's dispersion measure, DM$=11.6$ cm$^{-3}$ pc, corresponds to
$N_{\rm H} =3.6\times 10^{20}$ cm$^{-2}$ (assuming a 10\% ISM
ionization). This value is close to the total Galactic absorption in
neutral hydrogen, $3.6\times 10^{20}$ cm$^{-2}$ (Dickey \& Lockman
1990) in the direction of J1537 ($l=19\fdg85$, $b=48\fdg34$), which
is located well above the Galactic plane ($z=0.76$ kpc  at the
distance of 1 kpc). Therefore, we adopt $N_{\rm H}=3.6\times
10^{20}$ cm$^{-2}$ in the X-ray fits.

\begin{figure}
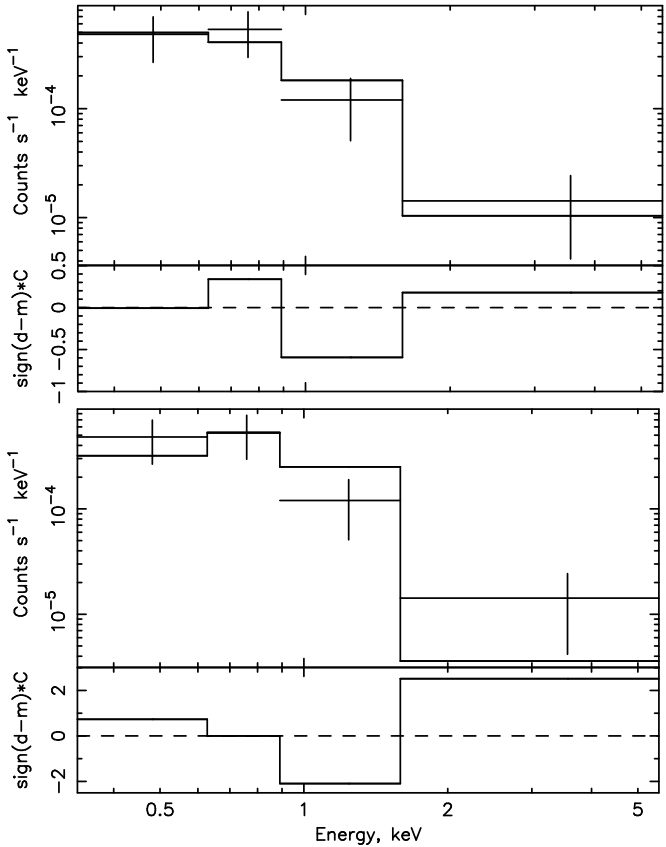

 \centering
 \vbox{
\includegraphics[width=2.1in,angle=-90]{f3a.eps}
\vspace{0.3cm}
\includegraphics[width=2.3in,angle=-90]{f3b.eps}}
\caption{ X-ray spectrum of J1537 fitted with PL ({\em top}) and BB
({\em bottom}) models. The contributions of the energy bins into the
best-fit C-statistic are shown in lower panels, multiplied by $-1$
when the number of data counts is smaller than the number of model
counts. }
\end{figure}

 The spectrum fits
a soft PL with a photon index
$\Gamma=
2.7$--3.7 and an isotropic unabsorbed 0.2--10 keV
 luminosity $L_{\rm X} \equiv 4\pi D^2 F_{\rm X}^{\rm unabs}
\simeq(6.1^{+3.0}_{-2.1})
\times 10^{29}$ ergs s$^{-1}$
 at
 $D =1$ kpc.
The temperature and the projected area of the emitting region
obtained from the BB fit are strongly correlated (see Fig.\ 4),
which results in large uncertainty of these parameters. The BB model
gives the
 best-fit temperature $T_{\rm BB} = 2.2$ MK and a projected emitting area
 $A= 5\times 10^{3}$ m$^2$. The projected area is much smaller than
 that of a NS, $\sim 3 \times10^{8}$ m$^{2}$, suggesting that the
 thermal radiation might be emitted from small, hot polar caps with an
 effective radius $R \sim 39 \langle \cos\varsigma\rangle^{-1/2}$ m and a
bolometric luminosity $L_{\rm bol}= \sigma T^4 A \langle
\cos\varsigma\rangle^{-1}=(6.8\pm2.1)\times10^{28}\langle
\cos\varsigma\rangle^{-1}$ ergs s$^{-1}$, where $\langle
\cos\varsigma\rangle$ is the time-averaged cosine of the angle between
the line of sight and the
normal to the visible polar cap ($\langle\cos\varsigma\rangle\approx 0.62$
 for the axis orientations suggested by Stairs et
al.\ 2004, assuming a centered dipole with two equivalent polar caps
and neglecting the GR bending of photon trajectories). The X-ray
luminosity of J1537 corresponds to $3.4\times10^{-4}\dot{E}_A$ and
$3.8\times10^{-5}\langle \cos\varsigma\rangle^{-1}\dot{E}_A =
6.1\times 10^{-5}\dot{E}$ for the PL (0.2--10 keV band) and BB fits,
respectively. The scarce statistics does not allow us to
discriminate  between the  BB and PL models.

To compare the spectra and light curves of J1537 and J0737,
we reanalyzed the {\sl
Chandra} ACIS and {\sl XMM-Newton} EPIC observations of J0737, using
the same spectral models and the same approach to the data analysis
as for J1537. The observations have been described by McLaughlin et
al.\ (2004), Campana et al.\ (2004), and Pellizzoni et al.\ (2004).
The results of our fits are presented in Table 3 and Figure 5. For
the PL fit of the ACIS spectrum, we obtained about the same
parameters as McLaughlin et al.\ (these authors did not fit the BB
model). As for the EPIC data, we found, in accordance with Campana
et al., that the EPIC-pn data are too noisy for a useful analysis because of
the very high background in the timing mode. Therefore, we only used
the MOS1 and MOS2 data.
For each of
the MOS detectors, we created a light curve with binning time of 10 s
and constructed GTIs excluding the time intervals during which the
total chip count rate exceeded 3 counts s$^{-1}$ (the remaining
useful time is 24.62 and 24.60 ks for MOS1 and MOS2, respectively).
The spectra we extracted from circular apertures
of $46''$ radius ($\simeq 90\%$ encircled energy). Comparing the properties
of the X-ray radiation from J0737 and J1537, we see that the
spectral parameters are very similar: $\Gamma \approx 3$ for the PL fit,
and $kT\approx 0.2$ keV for the BB fit.

\begin{figure}[]
 \centering
 \vbox{
\includegraphics[width=2.5in,angle=90]{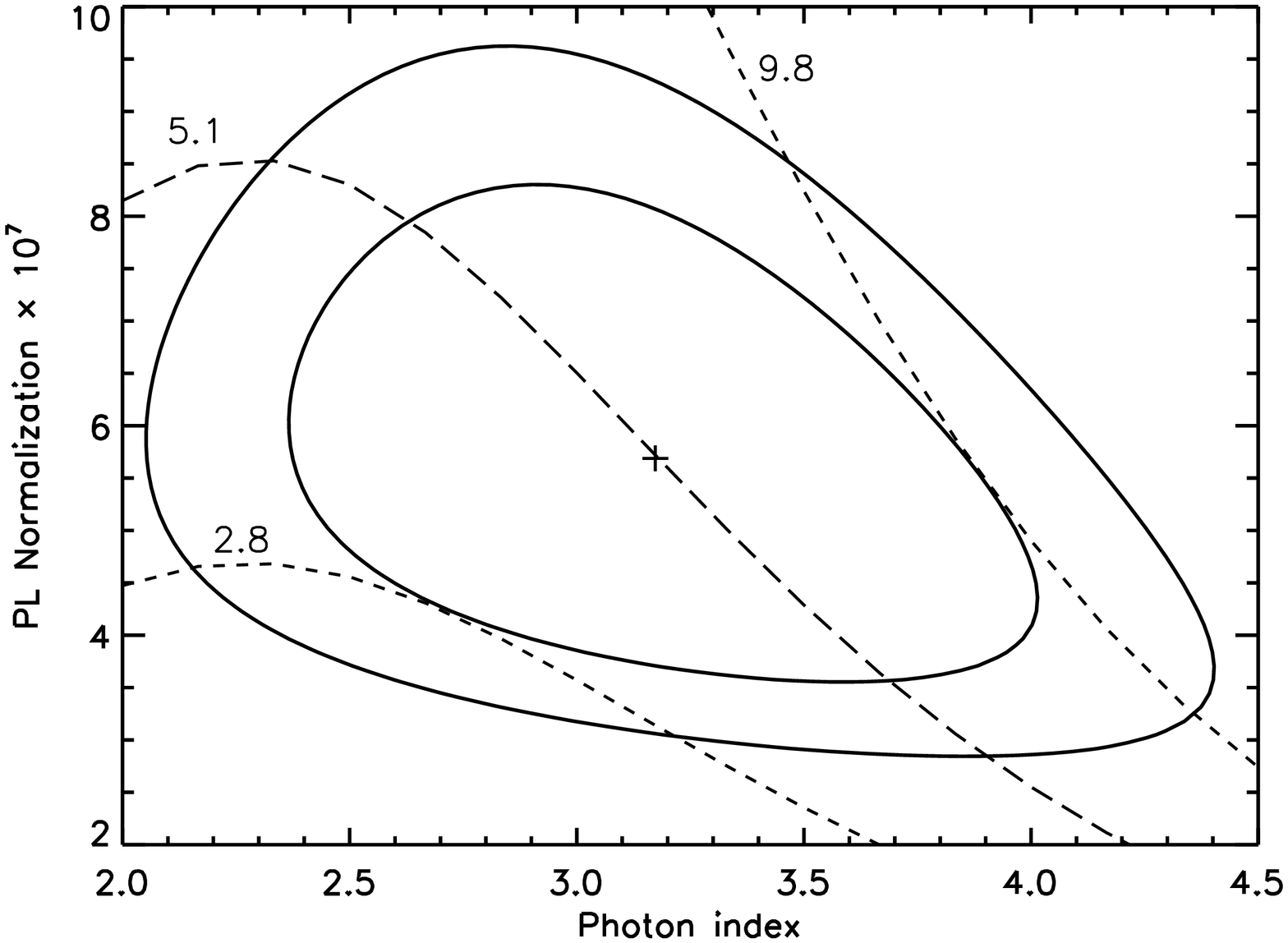}
\vspace{-0.0cm}
\includegraphics[width=2.5in,angle=90]{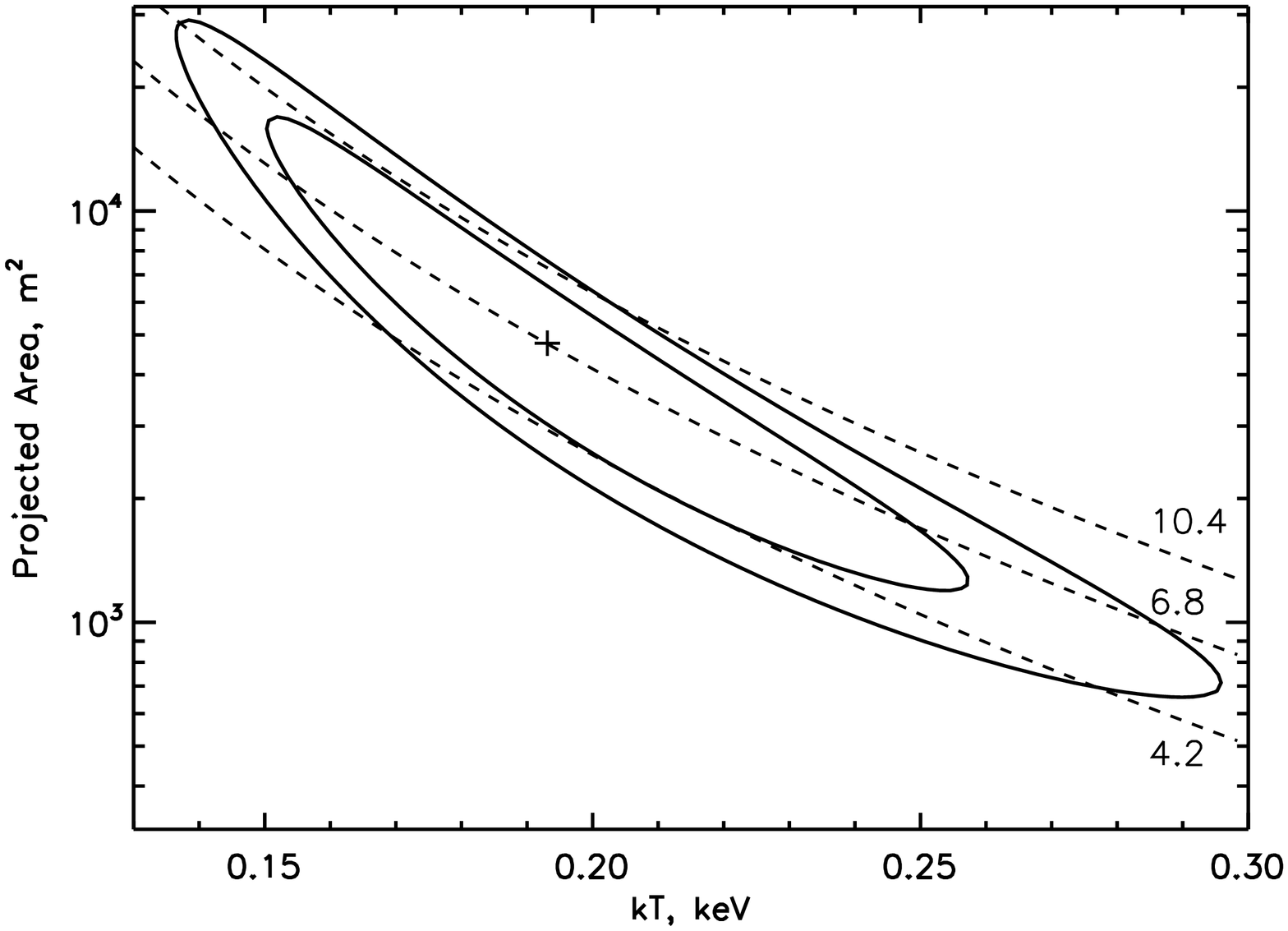}}
\caption{ Confidence contours (68\% and 90\%) for the PL ({\em top})
and BB ({\em bottom}) fits to the spectrum of J1537. The PL
normalization is in units of $10^{-7}$ photons cm$^{-2}$ s$^{-1}$
keV$^{-1}$ at 1 keV. The BB normalization (vertical axis) is the
projected emitting area in units of m$^2$, for $D = 1$ kpc. The
lines of constant unabsorbed flux (top panel; in units of 10$^{-15}$
ergs cm$^{-2}$ s$^{-1}$, in 0.2--10 keV band) and constant $L_{\rm
bol}\langle\cos\varsigma\rangle$ (bottom panel; in units of
10$^{28}$ ergs s$^{-1}$) are plotted as dashed lines, for fixed
$N_{\rm H} =3.6\times 10^{20}$ cm$^{-2}$.  }
\end{figure}

\begin{figure}[]
 \centering
 \vbox{
\includegraphics[width=2.5in,angle=90]{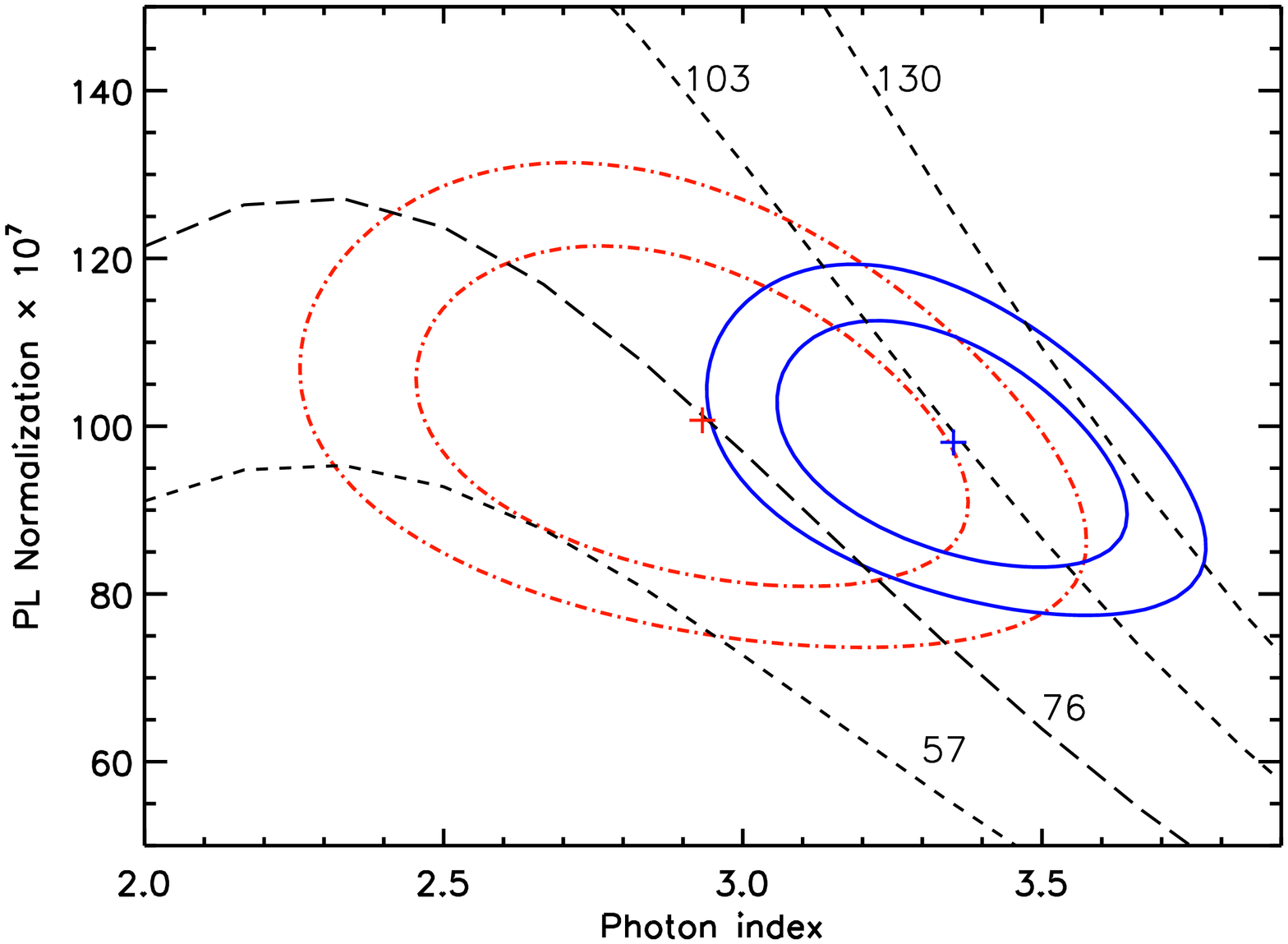}
\vspace{-0.0cm}
\includegraphics[width=2.5in,angle=90]{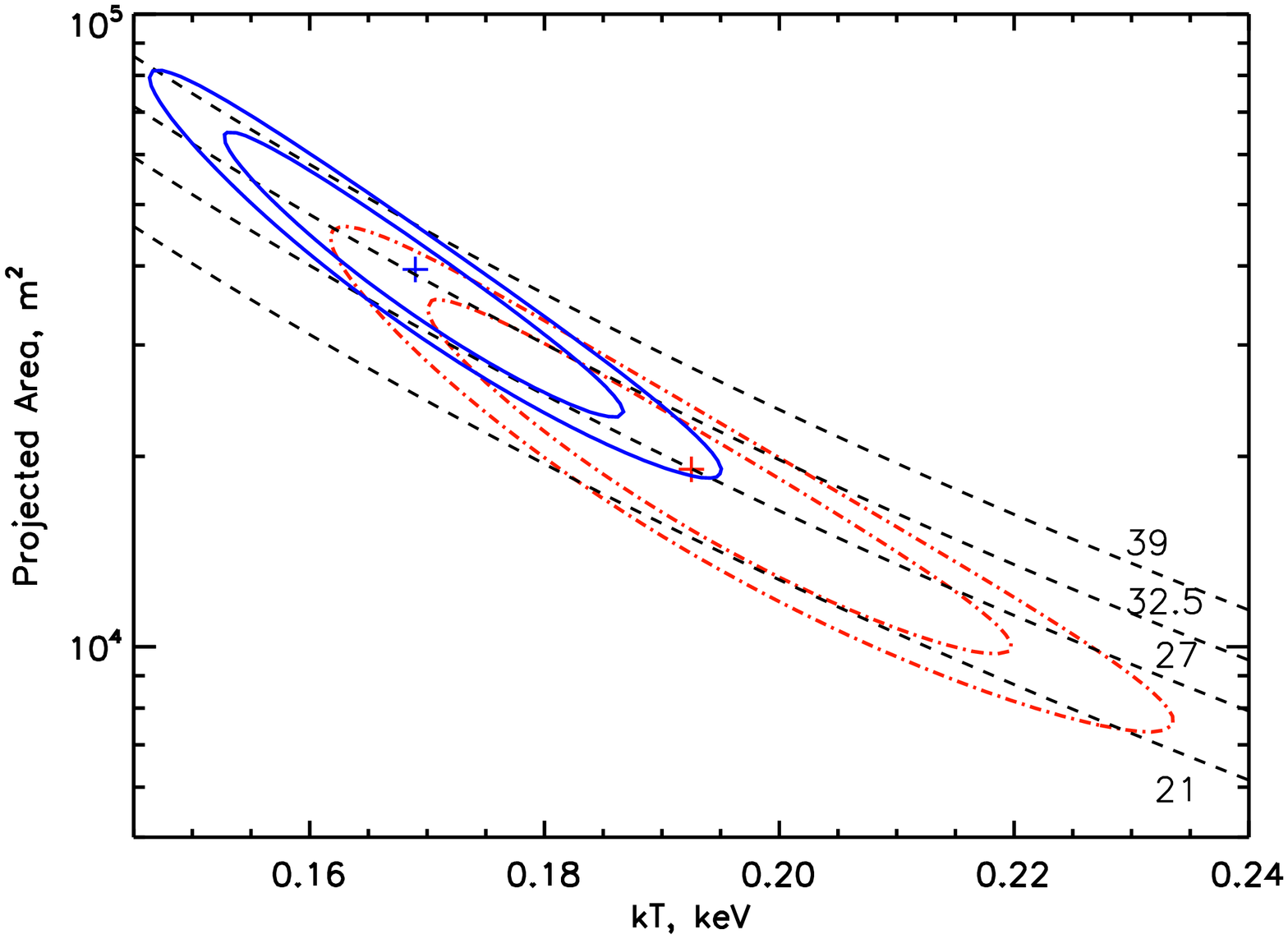}}
\caption{ Same as in Fig.\ 4 but for J0737. Solid and dash-dotted
contours are from the {\sl XMM-Newton} EPIC MOS and {\sl Chandra}
ACIS observations, respectively. {\em See the electronic edition of
the Journal for the color version of this figure.}}
\end{figure}

In the upper panel of Figure 6 we show the distribution of arrival
times for the 16 detected photons
over binary orbital phases of J1537
($\phi_{\rm orb} =0$ corresponds to
periastron passage)\footnote{To calculate the event orbital phases,
we used the binary ephemeris for the epoch of our observation
 kindly provided by Ingrid Stairs:
$P_{\rm orb} = 0.420737299122$ days, epoch of periastron
53470.309046790 MJD, longitude of periastron 290\fdg 004585816.}.
Notice that no events arrived during the 12,795 s interval between
MJD 53470.46046 and 53470.60855, which corresponds to the phase
interval [0.365; 0.717] ($\Delta\phi_{\rm orb} = 0.352$). This
interval includes the phase of apastron, $\phi=0.5$, and the phase
of superior conjunction of J0537A, $\phi = 0.408$ at the epoch of
our {\sl Chandra} observation. The probability of getting zero
counts by chance in the 12,795 s interval is 0.0034 (assuming
Poisson statistics), i.e.\ the observed ``gap'' has a
$\approx3\,\sigma$ significance. For comparison, we plotted similar
distributions for the \chan\ and {\sl XMM-Newton} observations of
J0737 (middle and bottom panels of Fig.\ 6), which do not show a
statistically significant dependence on orbital phase (Fig.\ 7).

\begin{figure}
 \vbox{
 \hspace{-1.0cm}
 \center
\includegraphics[width=2.5in,angle=90]{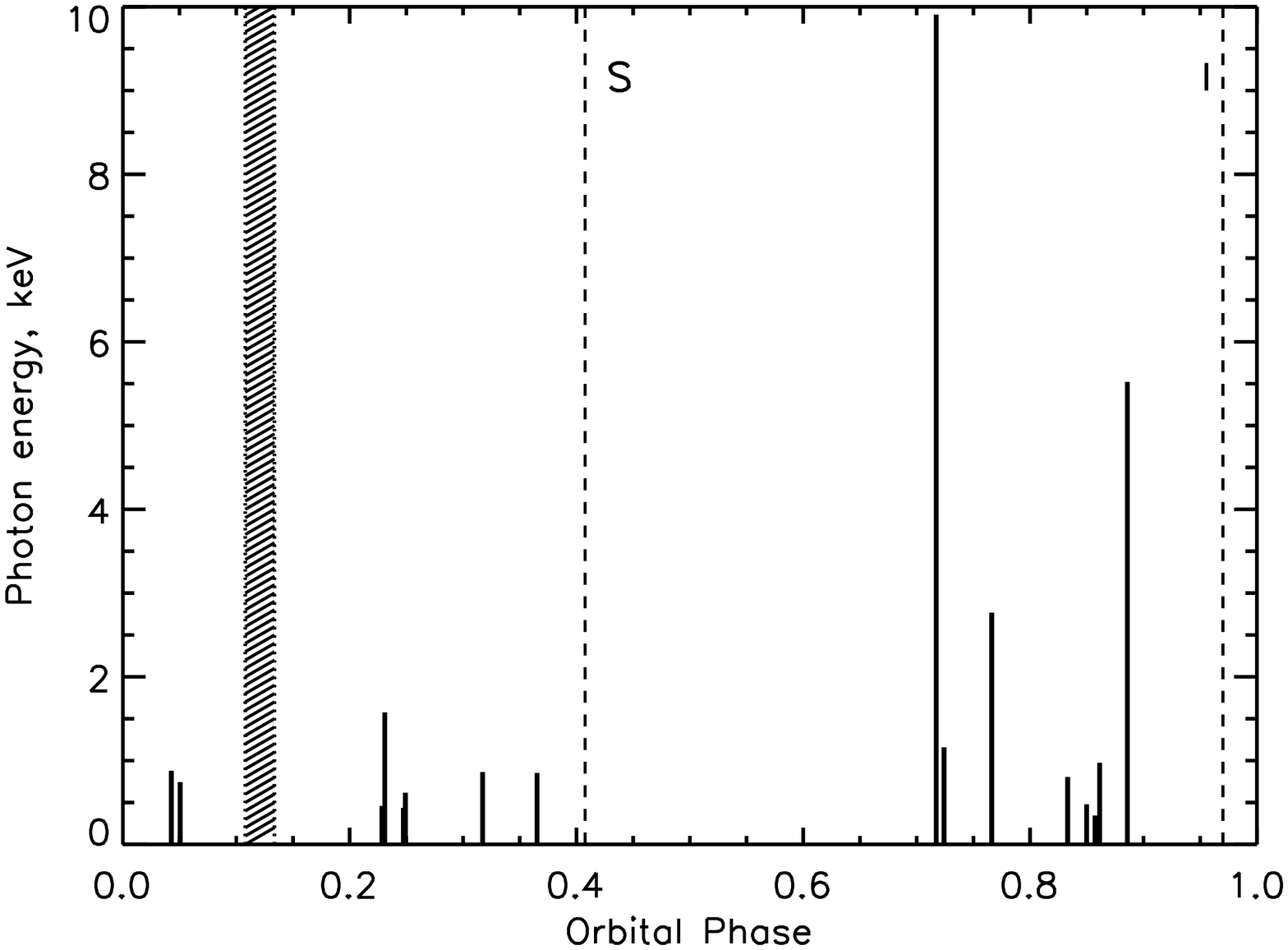}
\vspace{-0.3cm}
 %\hspace{-1.0cm}
\includegraphics[width=2.5in,angle=90]{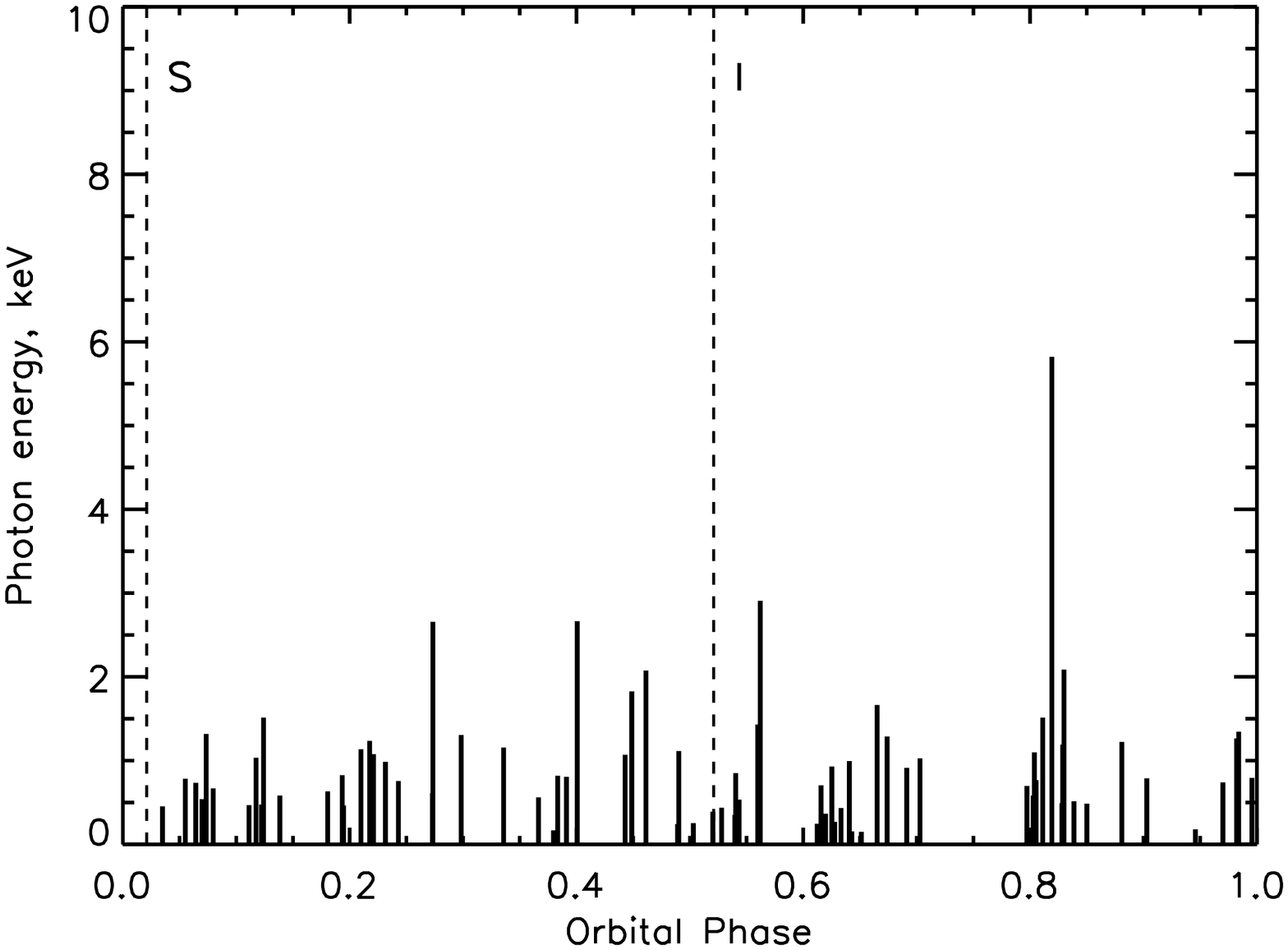}
\vspace{-0.3cm}
 %\hspace{-1.0cm}
\includegraphics[width=2.5in,angle=90]{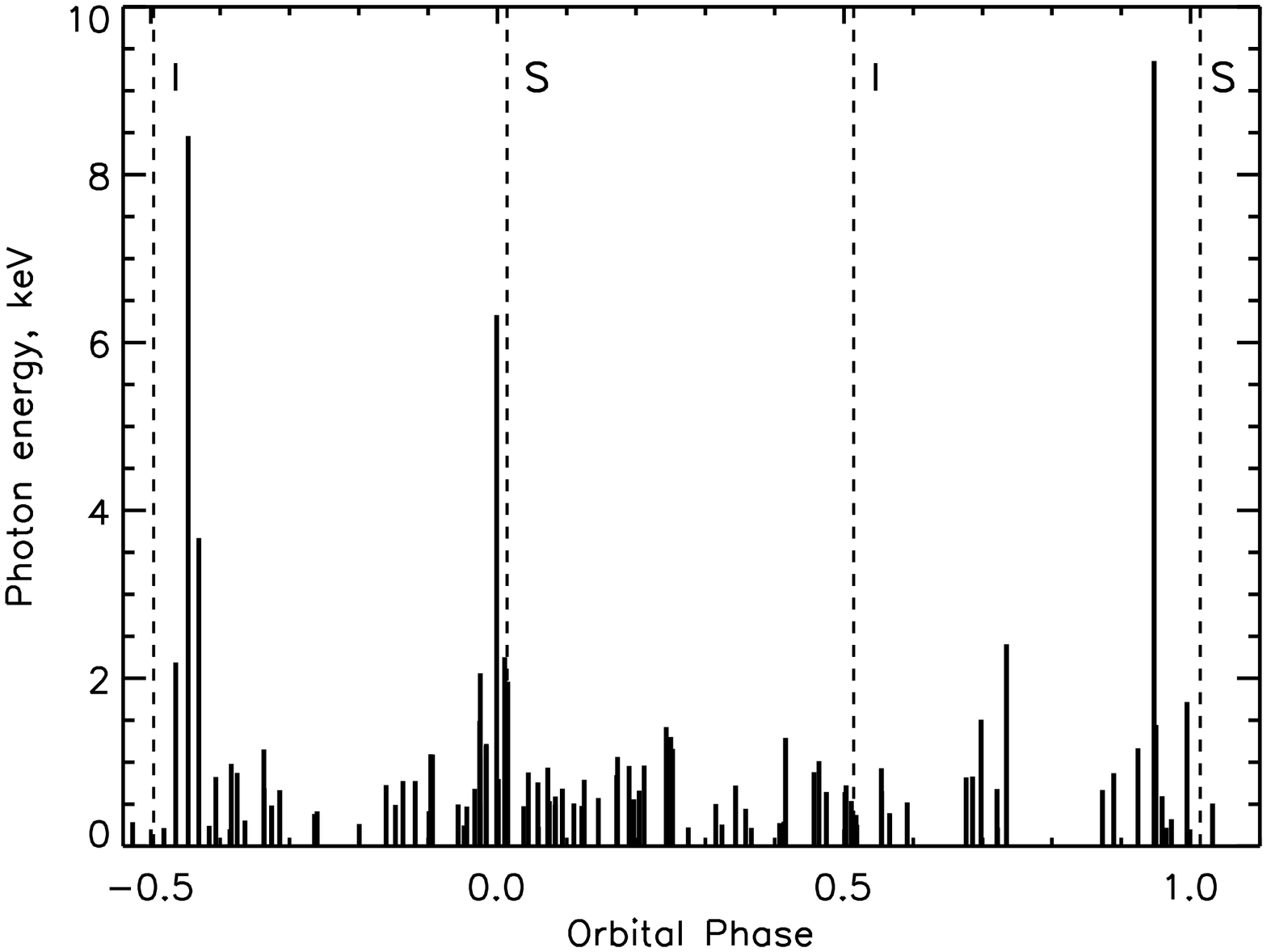}}
\caption{ Distributions of photon arrival times over binary phase
for the {\sl Chandra} ACIS observations of J1537 ({\em top}) and
J0737 ({\em middle}), and for the {\sl XMM-Newton} MOS1 and MOS2
observation of J0737 ({\em bottom}). Zero phase corresponds to
periastron passage. Only first 14.05 ks of the {\sl XMM-Newton}
exposure, least affected by the flare background, are used here. The
hatched area in the top panel shows the time interval excluded from
the GTIs. The dashed vertical lines show the phases of superior (S)
and inferior (I) conjunctions for J1537A and J0737A at the epochs of
the {\sl Chandra} and {\sl XMM-Newton} observations.
 }
\end{figure}

\begin{figure}
 \center
\includegraphics[width=2.5in,angle=90]{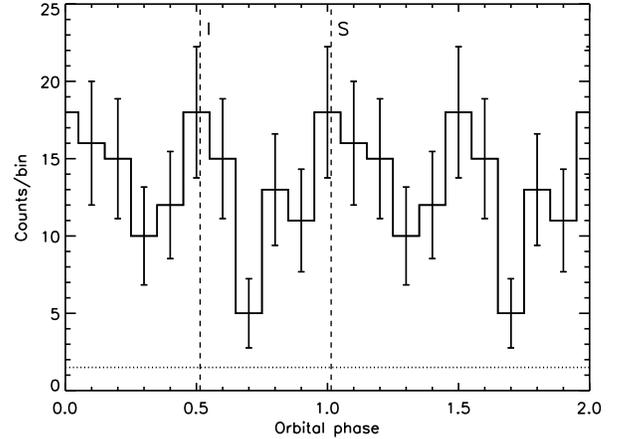}
\caption{ X-ray light curve of J0737 folded with the binary period.
We used both {\sl Chandra} ACIS and {\sl XMM-Newton} MOS data taken
over the length of a single binary revolution in each case (most of
the remaining 51-ks-long {\sl XMM-Newton} exposure suffered from
strong background flares.). The horizontal dotted line shows the
background level.
 }
\end{figure}

\begin{figure}
\centering
\includegraphics[width=3.4in,angle=0]{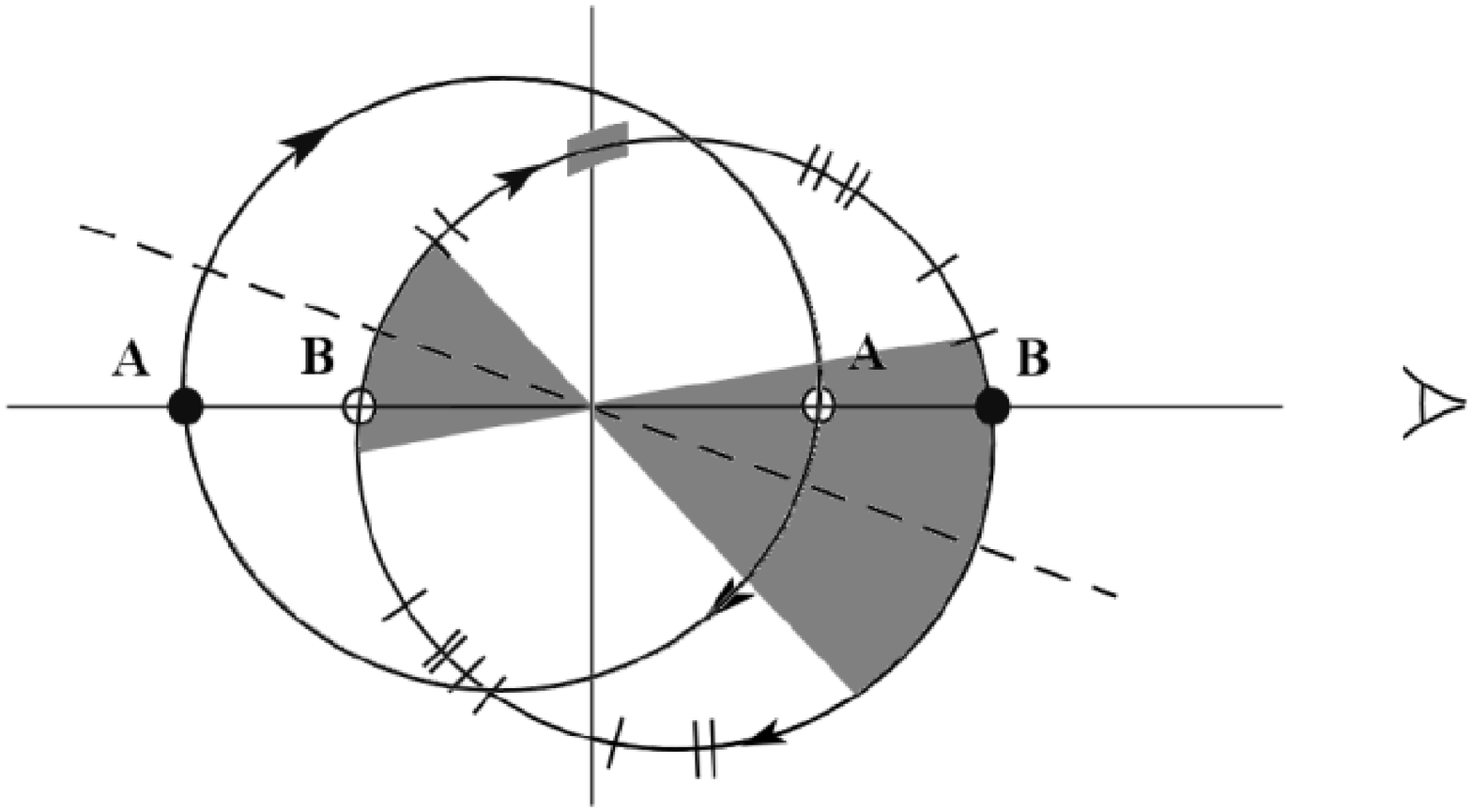}
\caption{Configuration of the J1537 binary at the epoch of the {\sl
Chandra} observation (53,470 MJD). View from above the orbital plane
with the Earth to the right is shown. The ellipses show the orbits
of J1537A and J1537B around the center of mass of the system. The
filled and open circles show the positions of the components at
superior and inferior conjunction of J1537A. The vertical solid line
is the line of nodes, and the dashed straight line is the apsidal
line. The ticks on the J1537B's orbit show the distribution of the
16 detected photons over the orbit, while the shaded part of the
orbit (around the line of nodes) corresponds to the excluded time
interval at the beginning of the observation. (Notice that the
distribution of photons over the orbit looks different from that
over phases in Fig.\ 6 because of the substantial eccentricity; in
particular, the gap near the periastron is much larger because of
the increased orbital velocity.) The shaded sectors show the ranges
of true anomalies $\theta$ (counted clockwise from periastron) in
which the lack of detected photons could be explained by a
misalignment of the A's spin axis and the binary angular momentum,
assuming an equatorial outflow of the A's wind (see \S4.1.1). }
\end{figure}

\section{Discussion.}

The small number of detected counts does not allow one to determine
the nature of the X-ray emission from J1537 unequivocally.
 Among viable models are intrinsic emission
from the surface or magnetosphere of the NS(s) (just as in
solitary pulsars), the synchrotron emission resulting from the
interaction between the pulsar winds or between the wind of J1537A
and the magnetosphere of J1537B, and
 emission from the
inner magnetosphere or surface of J1537B
triggered  by the relativistic particles from the J1537A's wind.
Below we briefly discuss these interpretations,
with emphasis on
different dependences of the X-ray flux on binary phase.

\subsection{
Phase-dependent X-ray emission powered by the wind
from pulsar A}

The
apparent lack of J1537's X-ray emission near the apastron suggests that a
phase-dependent interaction mechanism is responsible for the X-ray
emission in close DNSBs.
The interaction
 can proceed via several different
mechanisms, and the {\em pulsar winds} are expected to
play an important role in each of them. Below we consider
plausible
scenarios.

\subsubsection{
Interaction of pulsar A's wind with the outer magnetosphere of pulsar B}

As discussed in many papers on pulsars in binaries
(e.g., Arons \& Tavani 1993),
phase-dependent
synchrotron emission
can be
produced
when the
pulsar's wind collides with the wind or magnetosphere of
the secondary companion.
For a DNSB,
the position of the interaction site
(e.g., of the bow-shock head) on the line connecting
the two pulsars
can be estimated by balancing the
pressure of the pulsar A's wind with the pressure of pulsar B's
wind
 or magnetosphere:
\begin{eqnarray}
\frac{\dot{E}_{A} f_A}{4\pi c (d_{AB}-r_B)^2 c}
\approx\frac{\dot{E}_B}{4\pi c r_B^2} \left\{ \begin{array}{ll}
   f_B  & \mbox{if $r_B \gg R_{lc,B}$}\\
   g_B \left(\frac{R_{lc,B}}{r_B}\right)^{4} & \mbox{if $r_B \ll R_{lc,B}$}
  \end{array} ,
   \right.
\end{eqnarray}
where  $d_{AB}$ is the distance between the binary
companions, $r_B$ is the distance between the interaction site and
pulsar B, and $R_{lc,B}=4.8\times 10^9 P_B$ cm is the light
cylinder radius for pulsar B (approximately equal to
the size of undisturbed magnetosphere).
The factors $f_A$
and $f_B$
take into account possible
anisotropy of the pulsar winds
($f=1$ for an isotropic wind). They depend on the
colatitudes,
$\vartheta_A$ and $\vartheta_B$,
of the wind interaction site with respect to the A's and B's
spin axes  because the pulsar wind should be
axially symmetric;
for instance,
$f = (3/2) \sin^2\vartheta$ for an angular distribution
similar to that of magnetic dipole radiation
(cf.\ Arons et al.\ 2004), or
\begin{eqnarray}
 f =\left\{ \begin{array}{ll}
          (\sin \Delta)^{-1} & \quad \mbox{if $|\vartheta -\pi/2| < \Delta$}\\
          0  & \quad \mbox{otherwise}
          \end{array}
     \right.
\end{eqnarray}
for an equatorial outflow with a uniform energy flux
within sharp boundaries.
The factor $g_B=(3/4)(1+3\cos^2\Theta_B)$
takes into account the
dependence of
the magnetic pressure on the polar angle $\Theta_B$ with respect
to the B's magnetic axis.

Since B is expected to be
a slowly rotating, old (perhaps even dead) pulsar,
we can assume
$\dot{E}_{B}\ll \dot{E}_{A}$
(e.g., $\dot{E}_B/\dot{E}_A = 2.8\times
10^{-4}$ for J0737),
which gives
\begin{eqnarray}
 r_B \sim \left\{ \begin{array}{ll}
          d_{AB}(\dot{E}_{B}f_B/\dot{E}_{A}f_A)^{1/2} & \quad \mbox{for $r_B>R_{lc,B}$}\\
          d_{AB}^{1/3}R_{lc,B}^{2/3}(\dot{E}_{B}g_B /\dot{E}_{A}f_A)^{1/6}  & \quad \mbox{for $r_B<R_{lc,B}$}
          \end{array}  .
     \right.
\end{eqnarray}
The interaction may occur outside or
inside the B's light cylinder, depending on the separation
and  the unknown properties of
pulsar B. In particular, $r_B < R_{lc,B}$
(i.e., the B's magnetosphere is compressed by the A's wind
on the ``dayside'')
if
\be
P_B > 0.66\, d_{AB,11} (\dot{E}_{B,30} g_B/\dot{E}_{A,33} f_A)^{1/2}\, {\rm s},
\ee
which corresponds to
\begin{eqnarray}
P_B > \left\{ \begin{array}{ll}
0.3\, (g_B/f_A)^{1/2}\,{\rm s} & \quad \mbox{for J0737}\\
1.1\,(d_{AB}/a) (\dot{E}_{B,30} g_B/f_A)^{1/2}\,{\rm s} & \quad \mbox{for J1537}
\end{array}\, ,
     \right.
\end{eqnarray}
where
$a$ is the semimajor axis of the binary relative orbit
($a=2.28\times 10^{11}$ cm and $8.78\times 10^{10}$ cm for J1537
and J0737, respectively).

The shocked A's wind can generate
synchrotron X-ray emission if the wind particles
are sufficiently energetic and the magnetic field
$B$ is strong
enough in the emission region:
\be
\gamma^2 B \gtrsim
10^{12} (E/5\,{\rm keV})\,,
\ee
where $\gamma$
is the Lorentz factor,
and $E$ is the photon energy.
For the plausible case $r_B < R_{lc,B}$, the magnetic field at
the interaction site can be estimated from the pressure balance
between the A's wind and the B's magnetic pressure:
\begin{eqnarray}
B =
\left(\frac{2\dot{E}_Af_A}{c d_{AB}^2}\right)^{1/2}
=\frac{a}{d_{AB}} f_A^{1/2} \left\{ \begin{array}{ll}
7.1\, {\rm G} & \quad \mbox{for J0737}\\
1.5\, {\rm G} & \quad \mbox{for J1537}
\end{array} \, .
     \right.
\end{eqnarray}
At such fields, we need electrons with Lorentz
factors $\gtrsim 10^5$--$10^6$ to produce synchrotron X-rays.
The recent pulsar wind models (e.g., Kirk \& Skj{\ae}raasen 2003 and
references therein)
suggest that the initially
Poynting-flux-dominated wind of pulsar A
 converts most of its electromagnetic energy into particle
kinetic energy at a distance
much greater than the binary separation, which implies that
the bulk Lorentz factor upstream of the interaction site
remains the same as acquired in the pulsar A's magnetosphere.
Although the predictions of the current pulsar models
are rather uncertain in this regard (in particular,
the Lorentz factor depends on poorly known ``multiplicity''
of the pair cascades; e.g., Hibschman \& Arons 2001),
the required values of $\gamma$ are likely too large
to be produced in the A's magnetosphere, which means
that additional acceleration must occur at the interaction site
(e.g., caused by reconnection of the magnetic fields of the
A's wind and B's magnetosphere, similar to the magnetic reconnection
at the interaction of the solar wind with the Earth's magnetosphere).

If the shocked/accelerated A's wind is the sole source of the X-ray emission,
the X-ray luminosity should be
a fraction of the A's wind power:
\be
L_{\rm X} = \epsilon_\Omega \epsilon_e \epsilon_{\rm rad} \epsilon_{\rm X} \dot{E}_A\,
\ee
where $\epsilon_\Omega\approx f_A (r_\perp/2d_{AB})^2$ is the
fraction of A's power intercepted by the B's magnetosphere
with a transverse size $r_\perp$ that depends on geometry
of the interaction region (crudely, $r_\perp \sim R_{lc,B}$),
$\epsilon_e$ is the fraction of energy of the shocked/accelerated
wind carried by particles (electrons and/or positrons),
$\epsilon_{\rm rad} \approx {\rm min}(1;t_{\rm flow}/t_{\rm rad})$
is the radiative efficiency, i.e.
the fraction of intercepted energy emitted as (synchrotron)
radiation ($t_{\rm rad}$ and $t_{\rm flow}$ are the time of
synchrotron losses and the flow time in the emission region),
 and $\epsilon_{\rm X}$ is the fraction of synchrotron power
emitted in the X-ray range. The intercepted fraction can be
estimated as $\epsilon_\Omega \sim 6\times 10^{-3} f_A
(r_\perp/R_{lc,B})^2 $ and $\sim 5\times 10^{-4} f_A
(r_\perp/10^{10}\,{\rm cm})^2 (a/d_{AB})^2$ for J0737 and J1537,
respectively. The fraction $\epsilon_e$ depends on acceleration
mechanism (e.g., magnetic reconnection); the latter is currently
unknown, but the assumption that $\epsilon_e$ is not much lower than
unity seems plausible. The fraction $\epsilon_{\rm rad}$ depends on
the particle energy spectrum, geometry of emitting region, and flow
velocity. For a particle of given energy, the time of synchrotron
losses is $\tau_{\rm rad} = 510\, \gamma_6^{-1} B^{-2}\, {\rm s}$,
while the flow time is $t_{\rm flow} \sim l/v_{\rm flow} \sim 0.3\,
l_{10} \beta^{-1}\,{\rm s}$, where $l = 10^{10} l_{10}\,{\rm cm}$ is
the travel length, and $\beta = v_{\rm flow}/c$. This gives
$\epsilon_{\rm rad} \sim {\rm min}(1; 7\times 10^{-4} l_{10}\gamma_6
B^2 \beta^{-1})$. We see that electrons emit a very small fraction
of their energy unless the travel length substantially exceeds a
typical size of the B's magnetosphere. Finally, the fraction
$\epsilon_{\rm X}$ cannot be reliably estimated without knowing the
particle energy distribution. However, even for $\epsilon_{\rm X}$
(and $\epsilon_e$) close to unity, the above estimates of
$\epsilon_\Omega$ and $\epsilon_{\rm rad}$ suggest the A's wind
should be substantially anisotropic ($f_A\gg 1$) and/or the emitting
particles should spend a long time ($\gtrsim 10^2\,{\rm s}$) in the
emitting region to explain the ratio $L_{\rm X}/\dot{E}_A \sim
3\times 10^{-4}$ observed\footnote{We should note that the $L_{\rm
X}$ estimates in \S3 are based on the assumption of isotropic X-ray
emission, which is not necessarily true; the actual $L_{\rm X}$ can
be lower or higher than estimated.}
   in J0737 and J1537.

The interaction mechanism and the properties of the A's wind can be
inferred from the analysis of the dependence of the X-ray emission
on orbital phase. The collision of the A's wind with the B's
magnetosphere is expected to form a bow-shock, convex toward pulsar
A, with the shocked A's wind flowing in a magnetosheath engulfing
the B's magnetosphere (Arons et al.\ 2004; Lyutikov 2004).
 It is tempting to explain the
X-rays observed from J0737 and J1537 as synchrotron emission from
the shocked A's wind in the magnetosheath (as discussed by Granot \&
M\'esz\'aros 2004). Such an explanation implies an orbital
dependence of the X-ray flux caused by the relativistic beaming
(Doppler boost), showing a minimum flux at the phase of inferior
conjunction of pulsar A, when the wind in the sheath is flowing away
from the observer, and a maximum flux around the superior
conjunction. For a reasonable flow speed, $v_{\rm flow} = \beta c$,
the Doppler boost modulation can be quite large, up to $F_{\rm
max}/F_{\rm min} = [(1+\beta)/(1-\beta)]^{\Gamma +2}$, especially at
the observed large photon index, $\Gamma \sim 3$. However, the
observed distributions of photons over orbital phase do not show a
statistically significant flux deficit at the phase of A's inferior
conjunction, in neither J0737 nor J1537 (see Figs.\ 6 and 7).
Therefore, such an explanation can be ruled out. It seems surprising
that no orbital modulation caused by the anisotropy of the flow in
the putative magnetosheath is seen (unless, of course, our
assumption, that the X-rays are produced by interaction of the A's
wind with the B's magnetosphere, is wrong). We can only speculate
that either the magnetosheath does not form because the A's wind is
strongly magnetized or the X-ray emitting particles accelerated by
magnetic reconnection do not flow into the magnetosheath.

Having ruled out the magnetosheath as the source of the observed
X-rays, we can, nevertheless, suggest two
explanations for the observed orbital dependence of X-ray emission
in J1537, such that they do not contradict to the lack of such
dependence in J0737. First, this difference may be connected with
the fact that J1537 has a substantial eccentricity while J0737 is on
an almost circular orbit. At apastron, the separation $d_{AB}$ is
larger than at periastron by a factor $(1+e)/(1-e)$, which is 1.755
for J1537, vs.\ 1.192 for J0737. The decrease of the magnetic field
and particle number density, $n_e$, at the interaction site with
increasing separation should lead to a lower X-ray luminosity at
apastron in eccentric DNSBs. Crudely, assuming $\epsilon_{\rm X}$
does not depend on $d_{AB}$ substantially,
the luminosity behaves as
 $L_{\rm X} \sim \epsilon_{\rm X} n_e V {\cal P}_{\rm syn} \propto
V d_{AB}^{-4}$,
where ${\cal P}_{\rm syn} \propto B^2 \propto
d_{AB}^{-2}$ is the total synchrotron power per electron , and
$V$
is the emission volume. The dependence of the latter on $d_{AB}$ is
rather uncertain, but it seems reasonable to assume
$V\propto r_B^3$. Under this assumption,
 we obtain $V\propto d_{AB}^3$ or $V\propto d_{AB}$ for
$r_B> R_{lc,B}$ or $r_B< R_{lc,B}$, respectively, so the increase of $V$
with $d_{AB}$
does not compensate for the decrease caused by decreasing $B$ and $n_e$.
In particular, $L_{\rm X} \propto d_{AB}^{-3}$
for $r_B< R_{lc,B}$, which means that we can expect
the luminosity at apastron to be a factor of 5.4 lower than
at periastron, for J1537.
Although the effect of varying binary separation can explain
the observed deficit around apastron in J1537,
we do not see an increased count rate at periastron,
predicted by this model;
however, we cannot rule it out because of the scarce statistics
of our data.

In addition to the varying binary separation,
the phase dependence of the X-ray flux in J1537 can be caused by
a misalignment between the binary orbital plane and the equatorial
plane of pulsar A, where most of the A's wind is expected to be
confined. In this case, one should expect brighter X-ray emission
at orbital phases when pulsar B plunges in the dense wind flowing in the
equatorial plane of  pulsar A.
For instance, if the equatorial outflow is confined
between two conical surfaces, within an angle
$\pm\Delta$ from the equatorial plane, and the A's spin axis
is inclined at an angle $\delta$ to the direction of the orbital
angular momentum,
then, at $\delta > \Delta$,
 the wind ``misses'' pulsar B in two segments of the orbit:
$|\psi| < \eta$ and $|\pi -\psi| < \eta$, where $\psi$ is
the azimuthal angle
counted from the
projection of the A's spin
onto the orbital plane, and $\eta$ is given by the equation
\be
\cot\eta = \tan\Delta\,
(\tan^2\delta -\tan^2\Delta)^{-1/2} (\cos\delta)^{-1}\,.
\ee
In other words, one should expect a photon deficit in two
phase intervals corresponding to two sectors symmetric with
respect to the A's spin projection onto the orbital plane.
If such a phase dependence is observed, then $\eta$
can be measured, and $\Delta$ can be found from
the equation
\be
\tan^2\Delta = \cot^2\eta\,\sin^2\delta\,(1+\cot^2\eta\,\cos^2\delta)^{-1}.
\ee
For J1537, we determined the true anomaly $\theta$ for each of the 16 events
and marked the events on the J1537B orbit in Figure 8. Because of the
higher (lower) NS velocities near
the periastron (apastron), a relatively
small phase segment
around $\phi =0$
with no photons detected, $-0.114 < \phi < 0.043$,
translates into a sector of true anomalies,
$-68^\circ < \theta < 28^\circ$,
whose width, $\Delta\theta_{\rm per} = 95^\circ$,
is even larger than
that of the photon-free sector at apastron, $\Delta\theta_{\rm ap}=80^\circ$
($150^\circ < \theta < 231^\circ$).
 If we interpret the photon deficit
in these sectors as
caused by the anisotropy of the wind
outflow, then, from the symmetry requirement, the sectors in which
the J1537A's wind misses J1537B
(shaded in Fig.\ 8) are
$-30^\circ < \theta < 28^\circ$
and
$150^\circ < \theta < 208^\circ$, i.e.,
the projection of the A's spin onto the orbital plane
 almost coincides with the apsidal line,
and $\eta \leq 29^\circ$.
Since $\delta = 25\degr\pm4\degr$ for J1537 (Stairs et al.\ 2002),
equation (10) gives a lower limit on the half-opening angle
of the equatorial outflow, $\Delta \geq 18^\circ$--$25^\circ$.
Being consistent with the observed phase dependence,
such an explanation
 is also attractive because it
 implies an increased particle and energy supply in
the two segments of orbit where the J1537A's wind encounters J1537B
($f_A$ can be as large as 2.4--3.2, according to eq.\ [2]), a higher
magnetic field at the interaction region (by a factor of
$f_A^{1/2}$), and, correspondingly, a smaller ratio $t_{\rm
flow}/t_{\rm rad}$ (by a factor of $f_A$). These consequences of the
wind anisotropy alleviate the energetic and geometric restrictions
inherent in the interpretation of the X-ray emission as caused by
the interaction of the A's wind with the B's magnetosphere. We
should emphasize, however, that the X-ray flux deficit around
periastron is not statistically significant, and this hypothesis can
be confirmed only by a deeper observation.
  For J0737,
the angle $\delta$ is rather uncertain  ($0 < \delta < 60\degr$;
Manchester et al.\ 2005), so one can assume that the lack of a clear
orbital phase dependence means that $\delta < \Delta$, i.e., the
J0737A's wind interacts with J0737B throughout the whole orbit.

In addition to the energetics and geometry (luminosity and its phase
dependence), some information on the interaction mechanism can be
obtained from the X-ray spectra. In this regard, we should note that
the spectra observed in J1537 and J0737
 are surprisingly soft
($\Gamma \sim 3$) in comparison with, e.g., the spectra
observed in pulsar wind nebulae ($\Gamma \sim 1$--2).
The models of particle acceleration by ultrarelativistic MHD shocks
also suggest much flatter spectra,
$\Gamma\simeq1.6$, corresponding to the slope $p\simeq2.2$
of the particle energy spectrum
(Achterberg et al.\ 2001). We can speculate, however, that
acceleration via relativistic magnetic reconnection might produce
softer spectra. Also, we do not exclude the possibility that
the observed spectra consist of two components: a hard PL component
from the shocked wind and a soft thermal component from the polar caps
of pulsar A (see \S4.2).

\subsubsection{
Emission from B's inner magnetosphere and/or surface
induced by captured A's wind}
There is a possibility that some fraction of the A's
wind
is captured by the B's magnetosphere
(e.g., similar to capturing solar wind in the Earth magnetosphere).
The captured wind can increase the magnetospheric pair density
or even precipitate onto the NS surface
(such a possibility has been discussed by Zhang \& Loeb 2004
to explain
the flaring radio emission of J0737B).
The captured relativistic particles lose a substantial fraction of their
energy to synchrotron and inverse Compton radiation, thereby increasing
the radiative efficiency, $\epsilon_{\rm rad}$,
 of the intercepted wind. The particles
traveling along the open field lines may reach the NS polar caps and
heat them to X-ray temperatures, up to a few million kelvins,
 so that we could see not only nonthermal but also thermal X-ray emission,
which might explain the softness of the X-ray spectra observed in
J0737 and J1537. The total luminosity from these emission mechanisms
is, of course, a small fraction of $\dot{E}_A$; it can be described
by an equation similar to equation (8) but with an additional factor
in its right-hand side, $\epsilon_{\rm cap}$, which is the fraction of
intercepted wind particles captured by the B's magnetosphere.
Similar to the emission produced by the wind interaction with the
outer magnetosphere (\S4.1.1), the X-ray luminosity should decrease
with increasing $d_{AB}$ (i.e., it should be lower near apastron)
and vanish during the phase intervals when pulsar B is far from the
equatorial plane of pulsar A\footnote{We should mention that such
orbital phase dependencies are expected not only for the nonthermal
but also for the thermal component because the
polar cap cooling time scales
are very short (about a few microseconds for $B\sim 10^{12}$ G;
e.g., Gil et al.\ 2003).}. In addition, the fraction $\epsilon_{\rm
cap}$ should depend on the angle between the B's magnetic axis and
the line connecting the two pulsars. This may lead to an additional
modulation of $L_{\rm X}$ with the orbital period and the period of
pulsar B
 unless the B's magnetic
axis is perpendicular to the orbital plane, as expected if the main
braking mechanism for pulsar B is its interaction with the A's
wind (Arons et al.\ 2004; Lyutikov 2004).
In addition, the X-ray emission caused by the captured wind
should show pulsations with the B's spin period.

There is also an interesting possibility that the wind's particles
that reach the B's surface and heat the polar caps can facilitate
the extraction of electrons or ions from the polar cap surface by
the pulsar's electric field, triggering the pair cascade in pulsar
B. In this case, the A's wind works as a catalyst of B's X-ray (and
possibly radio) emission, providing an additional luminosity
associated with the loss of the pulsar B's spin energy\footnote{If
such a mechanism works, we should expect that $\dot{E}_B$ (and
$\dot{P}_B$) are increased during the ``activation phases''.}. The
orbital phase dependence of this luminosity should be similar to
that of the above-discussed captured wind. In particular, even if B
is a dead pulsar when no wind blows on it (e.g., when it is far from
the A's equatorial plane), it may become ``resurrected'' on the
orbit segments where it captures the wind.

Quantitative predictions of the properties of the X-ray emission
induced by the captured wind are highly uncertain, depending on
the unknown wind properties,
 structure of the magnetic field near the interaction
region, etc. We can crudely estimate the number of A's wind particles
captured by B per unit time:
$\delta\dot{N}_{A}=\kappa_A\dot{N}_{{\rm GJ},A}
\epsilon_\Omega \epsilon_{\rm cap}$, where
$\dot{N}_{{\rm GJ},A} \sim
(3\dot{E}_Ac/2e^2)^{1/2}$
is
the Goldreich-Julian rate of particle ejection from the A's magnetosphere,
$\kappa_A$ the pair production multiplicity, and
$\epsilon_\Omega \sim f_A (r_B/2d_{AB})^2$.
Assuming $r_B < R_{lc,B}$, we obtain
$\delta \dot{N}_A \sim 2.2\times 10^{27} \kappa_A \epsilon_{\rm cap}
(a/d_{AB})^{4/3} P_B^{4/3} (f_A^2g_B \dot{E}_{B,30})^{1/3}\,{\rm s}^{-1}$
for J1537, and $\delta \dot{N}_A \sim 4.4\times 10^{28} \kappa_A
\epsilon_{\rm cap} (f_A^2 g_B)^{1/3}\,{\rm s}^{-1}$ for J0737.
If the captured wind is the only source of X-rays, the product
$\kappa_A \epsilon_{\rm cap}$ should be large enough to provide the
observed X-ray luminosity: e.g., $\kappa_A \epsilon_{\rm cap} >
3\times 10^3 (f_A^2 g_B)^{-1/3} P_{B}^{-4/3} \dot{E}_{B,30}^{-1/3}
(a/d_{AB})^{-4/3} (10^5 m_ec^2/\bar{\cal E})$ in J1537
 and $\kappa_A \epsilon_{\rm cap} >
0.8\times 10^3 (f_A^2 g_B)^{-1/3} (10^5 m_ec^2/\bar{\cal E})$ in J0737,
where $\bar{\cal E}$ is an average energy per captured particle
reradiated in X-rays (Zhang \& Loeb 2004 assume $\kappa_A \sim
10^6$, $\epsilon_{\rm cap} \sim 0.1$ in their estimates for J0737;
as we have mentioned above, at such large $\kappa_A$ no X-rays can be
expected from the wind that was not captured). For $\kappa_A \sim
10^2$ obtained in standard cascade theory, we have to invoke the
above-mentioned activation of pulsar B by the captured wind to
explain the observed energetics.

Note that a large value of $\kappa_A\epsilon_{\rm cap}$
is also required for $\delta \dot{N}_{A}$ to exceed
the intrinsic B's pair production rate,
$\dot{N}_{B}=\kappa_B\dot{N}_{{\rm GJ},B}=
4.4\times 10^{29}\dot{E}_{B,30}^{1/2}\kappa_{B}\,{\rm s}^{-1}$.
It occurs
when
$\kappa_A\epsilon_{\rm cap}/\kappa_B >
200\, (d_{AB}/a)^{4/3} P_B^{-4/3} (f_A^2g_B)^{-1/3} \dot{E}_{B,30}^{1/6}\, {\rm s}^{-1}$
and
$13\, (f_A^2 g_B)^{-1/3}$ for J1537 and J0737, respectively.

Thus, we can
conclude that energizing the B's magnetosphere or heating
its surface by the particles from the A's wind
can explain the observed dependence of the X-ray
flux on binary phase. However, it would require
a rather high pair production multiplicity
in pulsar A and considerable anisotropy of the A's wind.

\subsection{
X-ray emission from individual pulsars}

Since solitary pulsars are known to emit X-rays, at least some
fraction of the detected X-rays could come from one or both of the
DNSB components, even if the X-ray emission caused by the
interaction of the A's wind with the B's magnetosphere were
negligible. In particular, MSPs can exhibit both nonthermal
(magnetospheric) and thermal (polar cap) components in their X-ray
spectra, the latter dominating in MSPs with $\dot{E} \lesssim
10^{34}$ ergs s$^{-1}$. Typical thermal conversion efficiencies,
$\eta_{\rm th}\equiv L_{\rm bol}/\dot{E}$, are
(0.3--$3)\times10^{-4}$ (Zavlin 2006, and references therein). The
efficiencies inferred from the BB fits of the J1537 and J0737
spectra, $6.1\times 10^{-5}$ and $4.6\times
10^{-5}\langle\cos\varsigma\rangle^{-1} D_{500}^2$, respectively,
are within that range\footnote{Since the orientations of the axes is
poorly known for J0737A (Demorest et al.\ 2004; Manchester et al.\
2005), the value of $\langle\cos\varsigma\rangle$ is very
uncertain.}. The polar cap radii derived from the BB fits, are
smaller than the conventional PC radii [$r_{\rm pc} \equiv (2\pi
R_{NS}^{3}/cP)^{1/2} = 740$ m and 960 m, respectively), similar to
other pulsars (e.g., Zavin \& Pavlov 2004; Zavlin 2006). Thus, the
luminosities and the  soft spectra of J1537 and J0737 are consistent
with being emitted from polar caps of J1537A and J0737A, but of
course such emission would not show any dependence on binary phase.

The X-ray emission in J1537 might also originate
from J1537B if this NS is not a ``dead pulsar'', i.e.\ it is
still capable of producing $e^{+}e^{-}$ pairs
emitting
curvature and synchrotron radiation in the magnetosphere and heating
its polar caps. Indeed, the X-ray spectrum of J1537 resembles the
spectra of
 old
($\tau \sim 1$--10 Myr) solitary radio pulsars (e.g., PSR B0950+08;
Zavlin and Pavlov 2004). However, the J1537's luminosity (e.g.,
$L_X= 8.5\times 10^{28}$ ergs s$^{-1}$ in the 1--10 keV band, for
the PL fit) is rather high for such an old pulsar\footnote{Based on
the J1537's proper motion and distance from the Galactic plane,
Thorsett et al.\ (2005) argue that the SN explosion that produced
J1537B occurred $\gtrsim 10$ Myrs ago.}; it would either imply a
high spin-down luminosity, $\dot{E}_{\rm J1537B}\gtrsim 10^{32}$
ergs s$^{-1}$,
 or suggest that the X-ray efficiencies of old
pulsars strongly grow with decreasing $\dot{E}$,
which is hardly supported
by observations
 (cf.\ Fig.\ 5 in Kargaltsev
et al.\ 2006).
Moreover, such an interpretation is firmly excluded for J0737,
whose X-ray luminosity exceeds $\dot{E}_{\rm J0737B} = 1.6\times 10^{30}$
ergs s$^{-1}$.

Thus, if the observed J1537's emission originates from the
surface or magnetosphere of one of the binary companions, J1537A
is a much more viable candidate than J1537B.
The assumption that most of the observed X-ray emission is generated
at the polar caps (or magnetosphere) of J1537A  would be strongly
supported if X-ray pulsations at the J1537A period are found in future
observations.
An indirect support for this interpretation would also be provided by
detection of X-ray pulsations from J0737 at the
J0737A's period.
On the other hand, confirming the reduced count rate near
apastron would virtually rule out this interpretation
and put an interesting upper limit on the MSP luminosity.

\subsection{Emission from the A's wind interacting with the interstellar
medium}

The velocities of J1537 and J0737 in the plane of the sky,
$v_{\perp}
\simeq 120$ and $70$
km s$^{-1}$, respectively,
are large enough to assume that these DNSBs
are moving through the ISM
with supersonic speeds, i.e., $v >
c_s = 15\, (\mu/0.6)^{-1/2} T_4^{1/2}$ km s$^{-1}$,
where $\mu$ and $T=10^4 T_4$ K are the molecular weight and
temperature. At a supersonic speed, the pulsar's wind forms a
termination shock (TS) in the ISM. For an isotropic wind,
the TS acquires
a bullet-like shape (Bucciantini, Amato, \& Del Zanna 2005),
with a distance $R_{\rm h} \simeq (\dot{E}_A/4\pi c \rho v^2)^{1/2}
=4.0\times 10^{15} \dot{E}_{A,33}^{1/2} n^{-1/2} v_7^{-1}$ cm
between the pulsar and the bullet head
[here $n=\rho/(1.66\times 10^{-24}\,{\rm g})$ is
the ISM number density in atomic mass units per cm$^3$,
and $v_7$ is the pulsar speed in units of 100 km/s].
For large enough $\gamma^2 B$ in the shocked plasma
(see equation [6]),
it can generate X-rays,
as discussed by Granot \& M\'esz\'aros (2004) for J0737.

The interpretation of the observed X-rays as generated by the
A's wind interaction with the ISM rather than with the B's magnetosphere
(\S4.1) has two apparent advantages for explaining the energetics
(but not the putative dependence on binary phase in J1537):
a much larger fraction of the A's wind is intercepted
(e.g., $\epsilon_\Omega \sim 0.5$),
and, possibly, a larger fraction of the A's wind energy is
carried by particles ($\epsilon_e$ is closer to unity)
because a larger fraction of the wind's magnetic energy
can be converted into particle energy at such large distances,
$R_{\rm h}\gg d_{AB}$.
On the other hand, this interpretation may have its own
problems, even with the X-ray energetics.
First of all,
since the wind's magnetic field is much lower
at the TS than at the site of interaction
with the B's magnetosphere,
the synchrotron
power and radiative efficiency, $\epsilon_{\rm rad}$, are much lower,
which requires much higher particle energies for the radiation
to be emitted in the X-ray range.
Assuming that the magnetization parameter (i.e., the ratio of the Poynting
flux to the kinetic energy flux) is small, $\sigma \ll 1$, we can,
following Kennel \& Coroniti (1986),  estimate
the post-shock magnetic field at the shock head:
$B \sim 3\sigma^{1/2} (4\pi\rho v^2)^{1/2} =
14\, \sigma_{-2}^{1/2} n^{1/2}
v_7\,\, \mu{\rm G}$, where $\sigma_{-2} =\sigma/10^{-2}$.
For the synchrotron radiation in such a field to be emitted in the
X-ray range, we should have enough electrons in the energy range
$\gamma \sim (1$--$3)\times 10^8 \sigma_{-2}^{-1/4} n^{-1/4} v_7^{-1/2}$.
Such energies are too high to be acquired
in the pulsar
magnetosphere.
The wind electrons can be further accelerated
in the pre-shock wind
by the conversion of the electromagnetic
wind's energy into kinetic energy. However,
irrespective of the (poorly understood) conversion mechanism, the maximum energy
is limited by the value
$\gamma_{\rm max} = (e^2/2 m_ec^3) \dot{N}_{{\rm GJ},A} \kappa_A^{-1}$
(see Lyubarsky \& Kirk 2001),
which gives $\gamma_{\rm max} =0.9\times 10^8 \kappa_A^{-1}$ and
$1.6\times 10^8 \kappa_A^{-1}$ for J1537A and J0737A, respectively.
These limiting energies are sufficiently large only at
very small multiplicities ($\kappa \lesssim 1 $ for reasonable
$\sigma$, $n$, and $v$),
much lower that $\kappa \sim 100$ expected from the pulsar models.
Too low electron energies and magnetic
fields may be responsible for the fact that no TSs
around supersonically moving MSPs have been
convincingly detected in X-rays
(e.g., Zavlin et al.\ 2002), in contrast to
forward bow shocks seen in H$_\alpha$.

It is worth mentioning another problem with the wind-ISM interaction
as the source of X-ray emission from
close binary pulsars,
such as J1537 and J0737. The conventional model for the shock
(and the above consideration) assume that the pulsar's velocity
relative to ISM is constant. However, a binary pulsar rotates
around the center of mass, and the velocity of
this motion may even exceed the binary systemic velocity.
For instance, the A's orbital velocity varies between 150 and 263 km
s$^{-1}$ in J1537 and between 275 and 328 km s$^{-1}$ for J0737.
This means that both the absolute magnitude and the direction of the
A's velocity relative to ISM vary with the orbital period. Moreover,
using the pulsar's velocity relative to ISM in the above
consideration would result in TS changing its position and
orientation in the ISM (e.g., the TS head could appear {\it behind}
the moving binary during some fraction of the binary period). On the
other hand, the characteristic time of propagation of perturbations
in the ultrarelativistic wind,
$\gtrsim R_{\rm h}/c \sim 30$--100 hours, exceeds the
binary period, so that we do not expect the wind-ISM interaction
to create a rotating (or wobbling) TS ``bullet''. Nevertheless,
even if this interaction results in a TS whose global properties,
averaged over the orbital period, resemble those of a solitary
pulsar moving with a constant speed $v$, this structure should respond to
the perturbations caused by the fast binary motion, and these perturbations
should manifest themselves in periodic variations in the radiation
from the shocked plasma.
However, no such variations have been seen in the X-ray radiation from J0737.
If future X-ray observations of J0737 put more stringent limit on the orbital
variation,
it will be an additional argument against the interpretation
in terms of the wind-ISM interaction.

\section{Conclusions}
We have detected the DNSB J1537 in X-rays, analyzed its X-ray spectrum
and light curve,
and compared the properties of its X-ray emission
 with those
of the double pulsar J0737. Although the X-ray spectra and
luminosities of these systems are similar, J1537 shows a gap in its
light curve during an interval of 0.35 of orbital period around
apastron while J0737 does not show show a significant orbital phase
dependence. There are two viable interpretations of the X-ray
emission: it can be caused by interaction of the pulsar A's wind
with pulsar B (\S4.1), including the capture of the wind particles
(\S4.1.2), or it can be emission from the magnetosphere or, more
likely, hot polar caps of pulsar A ( similar to solitary MSPs with
$\dot{E}< 10^{34}$ ergs s$^{-1}$; \S4.2). The former interpretation
 can explain the putative orbital phase dependence in J1537 if
the A's wind is mostly concentrated in the equatorial plane
around the A's spin axis
inclined to the orbital plane. The phase dependence can also be
associated with the large eccentricity of J1537 (much larger
than in J0737) because the X-ray luminosity depends on binary
separation. On the contrary,
no orbital phase dependence is expected if the X-rays are produced
by pulsar A.

 Unfortunately, the scarce count statistics
does not allow us to discriminate between the different
interpretations. Further observations of both J1537 and J0737, which
have different orbital parameters and orientations with respect to
the observer, should provide a clue to the nature of the X-ray
radiation from DNSBs and the properties of pulsars and their winds.

If future observations of J1537
confirm the deficit of X-ray flux during a
substantial part of
orbit,
this would prove that the X-rays are produced via interaction between
the binary companions. If a similar deficit is observed at the
opposite part of the orbit,
it would strongly
support the hypothesis that the A's wind is concentrated
towards the equatorial plane, and it would allow
one to measure the opening angle of the equatorial outflow.
If, on the contrary, the X-ray flux
shows a maximum at periastron, it would mean that the binary separation
is the main cause of the variable X-ray flux.
For J0737, the dependence of the
X-ray flux on orbital phase should be much weaker due to
a smaller angle between the orbital angular momentum
and the A's spin or/and the much
smaller eccentricity.
Confirming the interaction with the B's outer magnetosphere as
the source of X-rays would provide strong evidence that
the A's wind is accelerated
up to energies $\gtrsim 10^6 m_ec^2$ at the interaction site, likely
by magnetic reconnection.
In such a scenario, one can expect
X-ray pulsations with B's period,
especially if the wind is captured by pulsar B. If the emission spectrum
is found to be softer at the pulsations maxima,
we would conclude that a fraction
of the captured A's wind is
channeled onto B's polar
caps, heating them to X-ray temperatures.
Detection of radiation caused by the capture of the A's wind
would put a stringent lower limit on pair multiplicities in
pulsar A.

If
future observations rule out
 any orbital phase dependence in the DNSBs
(including the flux deficit near apastron in J1537),
 but
pulsations at
the A's period are detected,
 then we will have to conclude that the X-ray emission is
produced by pulsar A itself, while the interaction of the A's wind
with pulsar B is insignificant.
This would
mean
that conversion of the
intercepted wind's power
into X-ray luminosity is inefficient, possibly because
the particles are not accelerated to high enough energies
(which results in small values of $\epsilon_{\rm X}$ and $\epsilon_{\rm rad}$
in eq.\ [8]).
In this case, studying the X-ray spectrum and pulsations would
better characterize pulsar A, allowing one to compare its
properties with those of ``ordinary'' MSPs, which have shorter
periods and larger characteristic ages.

Finally, it is quite plausible that the observed emission
consists of two components: a soft thermal component from pulsar A
and a hard nonthermal component from the A's wind interaction with
pulsar B, as suggested by the X-ray spectra of J0737 and J1537.
If this is the case, we expect to see periodic variations
with both the orbital and A's periods. Also, the relative contributions
of these two components are expected to vary with orbital phase.
The changes can be seen through phase-resolved spectroscopy
because the wind emission is expected to be substantially harder than
the MSP emission.

To conclude, further deep
observations of J1537 and J0737, and perhaps other DNSBs,
would be of great interest
as they will constrain both the properties of
pulsar
winds, including their interaction
with the
neutron star companions, and the properties of pulsars themselves.

\acknowledgements
We thank Ingrid Stairs for providing the updated ephemeris for
J1537 and Firoza Sutaria for the help with the analysis of the
{\sl XMM-Newton} data.
This research was supported by NASA grants NAG5-10865 and
NAS8-01128 and {\sl Chandra} award SV4-74018.

\clearpage

\begin{deluxetable}{lll}
\tabletypesize{\scriptsize} \tablewidth{0pt}
\setlength{\tabcolsep}{0.005in} \tablecaption{ Observed and derived
parameters for J1537+1155 and J0737-3039 DNSBs}

\tablehead{ \colhead{Parameter} & \colhead{J1537+1155} & \colhead{J0737-3039} } % \\
\startdata
Binary period, $P_{b}$ (days)......................... & 0.421 & 0.102 \\
Relative semimajor axis, $a$ ($10^{11}$ cm) .... & 2.28 & 0.878 \\
Eccentricity, $e$ ........................................ & 0.274 & 0.0877 \\
Distance, $d$ (kpc).................................... & 1.0 &  $\approx 0.5$ \\
Orbital inclination, $\sin i$ ......................... & 0.975 & 1.000  \\
Transverse velocity, $v_\perp$ (km/s)................ &120 &70\\
Mass, $M$ ($M_\odot$)......................................   & 2.68  & 2.59 \\

\enddata
\tablecomments{Based on the data from Stairs et al.\ (2002, 2004),
Konacki et al.\ (2003), Lyne et al.\ (2004), Manchester
et al.\ (2005), and Cole et al.\ (2005).
The transverse velocoty in the plane of the sky,
$v_{\perp}$, is uncorrected for the solar system motion. }
\end{deluxetable}

\begin{deluxetable}{llll}
\tabletypesize{\scriptsize} \tablewidth{0pt}
\setlength{\tabcolsep}{0.005in} \tablecaption{ Observed and derived
parameters for J1537+1155A, J0737-3039A and J0737-3039B pulsars.}
\tablehead{\colhead{Parameter} & \colhead{J1537+1155A} & \colhead{J0737-3039A} & \colhead{J0737-3039B} } % \\
\startdata
Spin period, $P$~(ms)........................... &37.90 &  22.70 &2773.46 \\
Period derivative, $\dot{P}$ ($10^{-17}$).............. &
0.242 & 0.174 & 88 \\
Dispersion measure, DM (cm$^{-3}$~pc)\,...
& 11.6 &  48.9 & 48.7\\
Angle $\delta$ (deg) ..................................... & $25\pm4$ (or $155\pm4$) &$0-60$ (or $120-180$) & ... \\
Angle $\alpha$ (deg) .................................... &
103 & $0-60$ & ... \\
Mass ($M_\odot$) .......................................  & 1.33 & 1.34 & 1.25 \\
Semimajor axis, $a$ ($10^{11}$ cm) .............. & 1.12  & 0.424 & 0.454 \\
 Surface magnetic field, $B_s$~($10^{10}$ G)....
& 0.97 &  0.64  & 158\\
Spin-down power, $\dot{E}$~($10^{33}$ erg s$^{-1}$)....
& 1.8 &  5.8 & 0.0016 \\
Age, $\tau=P/(2\dot{P})$, (Myr).....................
& 248 &  210  &  50 \\
Light cylinder radius, $R_{lc}$ ($10^{8}$ cm) ... & 1.8& 1.1 & 130 \\
Magnetic field at $R_{lc}$, $B_{lc}$ (G) ......... & 1660 & 5050 & 0.69 \\
Polar cap radius, $R_{\rm pc}$ (m) ..................& 740 &960 &87 \\
Primary pair flux, $\dot{N}_{\rm GJ}$ ($10^{30}$ s$^{-1}$)...... & 19 & 34 & 0.6\\

\enddata
\tablecomments{Based on the data from Stairs et al.\ (2002, 2004), Lyne et al.\ (2004),
and Manchester et al.\ (2005).
Angle $\delta$ is between the pulsar's spin and orbital
rotation axis. Angle $\alpha$ is between the pulsar's spin and
magnetic axis. The surface magnetic field, $B_s$, is assumed to be
that of a dipole and given at the equator. }
\end{deluxetable}

\begin{table}[]
\caption[]{Parameters of PL and BB models fitted to the {\sl
Chandra} and {XMM-Newton} spectra of J1537 and J0737}
\vspace{-0.5cm}
\begin{center}
\begin{tabular}{cccccccc}
\tableline\tableline Instrument & Model &
$N_{H,20}$\tablenotemark{a}&Norm.\tablenotemark{b} &
$\Gamma$ or $kT$\tablenotemark{c} & C/dof & Q\tablenotemark{d} & $L_{\rm X}$ or $L_{\rm bol}\langle\cos\varsigma\rangle$\tablenotemark{e} \\
\tableline
& ~~~~ & ~~~ & ~~~~~~~~~~~{\bf J1537+1155} & & \\
    ACIS & PL          &       $3.6$&$5.7^{+1.6}_{-1.5}\times 10^{-7}$ &  $3.17^{+0.52}_{-0.52}$ & 1.1/2    & 12\% & $6.1^{+3.0}_{-2.0}\times 10^{29}$  \\
    ACIS  & BB         &       $3.6$&$4.8^{+5.5}_{-3.0}\times 10^{7}$  &  $193_{-30}^{+37}$      &  $5.3/2$ & 76\% & $6.8^{+2.1}_{-2.1} \times 10^{28}$   \\
\tableline
 & ~~~~ & ~~~ & ~~~~~~~~~~~~{\bf J0737--3039}
            & & \\
   ACIS  & PL          &       $5.0$&$1.01^{+0.14}_{-0.16}\times10^{-5}$   &  $2.93^{+0.29}_{-0.31}$ & 1.7/4    & 5\%  & $2.27_{-0.39}^{+0.42}\times 10^{30}$  \\
   ACIS  & BB          &       $5.0$&$1.91_{-0.71}^{+0.99}\times 10^8$     &  $192^{+17}_{-15}$      & 5.7/4    & 52\% & $2.67^{+0.33}_{-0.37}\times 10^{29} $   \\
   MOS1+2  & PL        &       $5.0$&$0.98^{+0.08}_{-0.10}\times10^{-5}$   &  $3.35^{+0.19}_{-0.19}$ & 10.0/14  & 12\% & $3.03_{-0.49}^{+0.50}\times 10^{30}$  \\
   MOS1+2  & BB        &       $5.0$&$3.95_{-1.0}^{+1.5}\times 10^8$       &  $169^{+11}_{-13}$      & 16.9/14  & 62\% & $3.28^{+0.40}_{-0.38}\times 10^{29} $   \\
\tableline
\end{tabular}
\end{center}
\tablecomments{ The errors are at the 68\% confidence level for one
interesting parameter. } \tablenotetext{a}{Fixed values of hydrogen
column density, in units of $10^{20}$ cm$^{-2}$.}
\tablenotetext{b}{Normalization: Spectral flux in photons cm$^{-2}$
s$^{-1}$ keV$^{-1}$ at 1 keV or projected area of the emitting
region in cm$^2$, for the PL and BB models, respectively.}
\tablenotetext{c}{Photon index or BB temperature in eV.}
\tablenotetext{d}{Percentage of 10,000 Monte Carlo simulations,
drawn from the best-fit model, which give a C-statistic value lower
than the best-fit value. Although the BB fits are formally worse
than the PL fits, they are statistically acceptable. }
\tablenotetext{e}{ $L_{\rm X}$ is the unabsorbed PL luminosity in
the 0.2--10 keV band, $L_{\rm bol}$ the bolometric luminosity for
the BB fits. }
\end{table}

\end{document}